\documentclass[utf8]{frontiersSCNS} 
\usepackage{url,hyperref,lineno,microtype,subcaption}
\usepackage[onehalfspacing]{setspace}

\def\keyFont{\fontsize{8}{11}\helveticabold }
\def\firstAuthorLast{Karampelas {et~al.}} 
\def\Authors{Konstantinos Karampelas\,$^{1,*}$, Tom Van Doorsselaere\,$^{1}$, David J. Pascoe\,$^{1}$, Mingzhe Guo\,$^{2,1}$ and Patrick Antolin\,$^{3}$}
\def\Address{
$^{1}$Centre for mathematical Plasma Astrophysics, Department of Mathematics, KU Leuven, Celestijnenlaan 200B bus 2400, 3001 Leuven, Belgium \\
$^{2}$Institute of space sciences, Shandong University, Weihai 264209, China\\
$^{3}$School of Mathematics and Statistics, University of St. Andrews, St. Andrews, Fife KY16 9SS, UK}
\def\corrAuthor{Konstantinos Karampelas}
\def\corrEmail{kostas.karampelas@kuleuven.be}

\begin{document}
\onecolumn
\firstpage{1}

\title[Simulated decayless oscillations]{Amplitudes and energy fluxes of simulated decayless kink oscillations.} 

\author[\firstAuthorLast ]{\Authors} 
\address{} 
\correspondance{} 

\extraAuth{}

\maketitle

\begin{abstract}
\section{}

Recent observations with the Atmospheric Imaging Assembly (AIA) instrument on the SDO spacecraft have revealed the existence of decayless coronal kink oscillations. These transverse oscillations are not connected to any external phenomena like flares or coronal mass ejections, and show significantly lower amplitudes than the externally excited decaying oscillations. Numerical studies have managed to reproduce such decayless oscillations in the form of footpoint driven standing waves in coronal loops, and to treat them as a possible mechanism for wave heating of the solar corona. Our aim is to investigate the correlation between the observed amplitudes of the oscillations and input the energy flux from different drivers. We perform 3D MHD simulations in single, straight, density-enhanced coronal flux tubes for different drivers, in the presence of gravity. Synthetic images at different spectral lines are constructed with the use of the FoMo code. The development of the Kelvin-Helmholtz instability leads to mixing of plasma between the flux tube and the hot corona. Once the KHI is fully developed, the  amplitudes of the decayless oscillations show only a weak correlation with the driver strength. We find that low amplitude decayless kink oscillations may correspond to significant energy fluxes of the order of the radiative losses for the Quiet Sun. A clear correlation between the input energy flux and the observed amplitudes from our synthetic imaging data cannot be established. Stronger drivers lead to higher vales of the line width estimated energy fluxes. Finally, estimations of the energy fluxes by spectroscopic data are affected by the LOS angle, favoring combined analysis of imaging and spectroscopic data for single oscillating loops.

\tiny
 \keyFont{ \section{Keywords:} solar corona, forward modeling, magnetohydrodynamics, corona loops, decayless oscillations}
\end{abstract}

\section{Introduction}

Over the past twenty years, observations of the Sun have shown the existence of waves and oscillations throughout the solar corona \citep{aschwanden2006,demoortel2012review}. The discovery of transverse magnetohydrodynamic (MHD) standing \citep{aschwanden1999, nakariakov1999} waves in coronal loops, and propagating waves in open magnetic field structures \citep{verwichte2005} has lead to many observational and numerical studies. The ubiquity of such waves has also been established in prominence threads \citep{okamoto2007}, coronal loops \citep{mcintosh2011}, as well as greater areas of the corona \citep{tomczyk2007, tomczyk2009, thurgood2014ApJ, morton2016ApJ}, renewing the interest on the effects of these waves in the solar atmosphere.  

Analytical studies on the nature of the transverse oscillations in inhomogeneous plasmas \citep{zajtsev1975, ryutov1976, edwin1983wave,allcock2017} have described the different surface waves expected in a non-uniform plasma. In order to explain the observed damping of such oscillations \citep{tomczyk2009,verthterradas2010,terradas2010, Pascoe2016A&Aobs, Pascoe2017A&A,Pascoe2019Frontiers}, extensive theoretical and numerical work has also been performed. The mechanisms of resonant absorption and mode coupling \citep{sakurai1991, goossens1992resonant, ruderman2002damping, arregui2005resonantly, goossens2011resonant, pascoe2012, pascoe2016A&A, demoortel2016, yudaejung2017ApJ, Pascoe2018ApJ} are considered the reason behind this spatial and temporal attenuation of the oscillations, by transferring the energy of the global kink mode to local azimuthal Alfv\'{e}n modes. Through  phase mixing \citep{heyvaerts1983,soler2015}, the energy is then transferred to ever decreasing smaller scales until it gets dissipated by resistivity and viscosity \citep{ofman1994heat,ofman1994bheatvis,poedts1996,ofman1998}. Observational studies of waves in the solar chromosphere and corona \citep{depontieu2007,tomczyk2009,Morton2012NatCo} suggest the existence of enough energy flux to sustain the radiative losses of $\sim 100$ W m$^{-2}$ for the non-active region corona \citep{withbroenoyes1977ara}. In \citet{demoortel2012ApJ} it was shown that LOS integration of footpoint driven multistrand coronal loop oscillations leads to an underestimation of the wave energy. In \citet{antolin2017} it was found for a non-driven oscillating loop that the wave energy ends up underestimated through the localisation of the energy by resonant absorption. Recent simulations of coronal loop waves \citep{pagano2017,pagano2018,Pagano2019arXivomf} have not reported sufficient heating to balance the radiative losses.

Alongside the high amplitude, externally initiated, decaying transverse oscillations in coronal loops \citep{verwichte2009ApJ, verwichte2010ApJ,WhiteVerwichte2012A&A,White2012A&A} a new group of small-amplitude decayless transverse oscillations have been identified in coronal loops \citep{nistico2013,anfinogentov2015,duckenfield2018}, with amplitudes $\sim 0.1$-$0.4$ Mm. These decayless oscillations have been interpreted in different ways over the years. They have been treated as continuously driven 
kink waves with a footpoint driver \citep{mingzhe2019,karampelas2019,afanasev2019}, as a self-oscillatory process due to
the interaction of the loops with quasi-steady flows \citep{nakariakov2016}, or as a line of sight (LOS) effect from the development of the Kelvin-Helmholtz instability (KHI) \citep{antolin2016} in impulsive standing loop oscillations.

The development of the KH instability has been theorized in plasma structures were standing surface waves are observed \citep{heyvaerts1983,BrowningPriest1984,zaqarashvili2015ApJ, Barbulescu2019ApJ, Hillier2019MNRAS}, caused by the strong shear velocities generated by the azimuthal Alfv\'{e}n waves. Recent numerical studies \citep{terradas2008,antolin2014fine, magyar2016damping,howson2017twisted,terradas2018,karampelas2017, antolin2018ApJ...856...44A} have confirmed the development of transverse wave induced Kelvin-Helmholtz (TWIKH) rolls for standing kink waves. Additional work has been performed in order to develop methods of identifying the effects of KHI in oscillating loops \citep{Goddard2018ApJ,tvd2018DEM}. Spatially extended TWIKH rolls have been found in the case of continuously driven standing waves, which fully deform the initial monolithic loop cross-section into a fully turbulent one \citep{karampelas2018fd,karampelas2019}. These spatially extended TWIKH rolls in simulations of continuously driven loops have also been reported as sites of mixing of plasma and heating in the solar corona, by effectively spreading the effects of phase mixing across the cross-section of loops \citep{karampelas2017,mingzhe2019,karampelas2019,afanasev2019}. 

One of the main challenges in creating an efficient wave heating model is to find a way of providing a high enough energy flux over large periods of time, while still remaining within the constraints imposed by the observational data. The decayless oscillations seem to be ideal candidates for prolonged energy input in such a model, and also pose the constraint of having relatively small oscillation amplitudes. In \citet{mingzhe2019}, the use of a mixed Alfv\'{e}n and transverse footpoint driver suggested as a way of increasing the total energy input while retaining small amplitudes in the synthetic images. The authors in \citet{afanasev2019} suggest the use of broadband drivers as a way to include additional energy for a decreased observational impact, while still mantaining sufficient energy input. In the current work, we study the effects of transverse footpoint drivers of different strengths on the observational signatures of the induced oscillations in synthetic images. The aim is to detect the existence of a correlation between the input energy flux and the oscillation amplitudes and to determine the observed fluxes from  synthetic spectroscopic and imaging data, and set possible constraints for future work.

\section{Material and Methods}

\subsection{Basic setup}

For our $3D$ simulations, we use straight, density-enhanced magnetic flux tubes in a low-$\beta$ coronal environment, following  \citet{karampelas2019}. This setup models gravitationally stratified, active region coronal loops in ideal MHD, in the presence of numerical resistivity. Each loop has a full length ($L$) of $200$ Mm and an initial minor radius ($R$) of $1$ Mm, which is constant with height. The radial density profile of our cylindrical flux tubes at the footpoint is
\begin{equation}
\rho(x,y) = \rho_e  + (\rho_i - \rho_e)\dfrac{1}{2}(1-\tanh((\sqrt{x^2+y^2}/R-1)\,b)).\label{eq:1}
\end{equation}
We denote the basic values of our physical parameters with the index $i$ ($e$) for internal (external) values, with respect to our tube. The external and internal density at the footpoint are equal to $\rho_e = 10^9 \mu \, m_p$ cm$^{-3} = 0.836 \times 10^{-12}$ kg m$^{-3}$ ($\mu = 0.5$ and $m_p$ is the proton mass) and $\rho_i = 3\times \rho_e$. We denote the coordinates in the plane perpendicular to the loop axis by $x$ and $y$ , and by $z$ the coordinate along its axis. $b=20$ sets the width of the boundary layer to $\ell \sim 0.3 R$. For the models studied, we set the temperature to be constant with height and equal to $T_e = 3\times T_i = 2.7$ MK. Finally, we consider an initial uniform magnetic field parallel to the flux tube axis (along the $z$-axis) equal to $B_z=22.8$ G. This gives us a plasma $\beta = 0.018$.

\begin{figure}[h!]
\begin{center}
\includegraphics[trim={0cm 0cm 0cm 0cm},clip,scale=1]{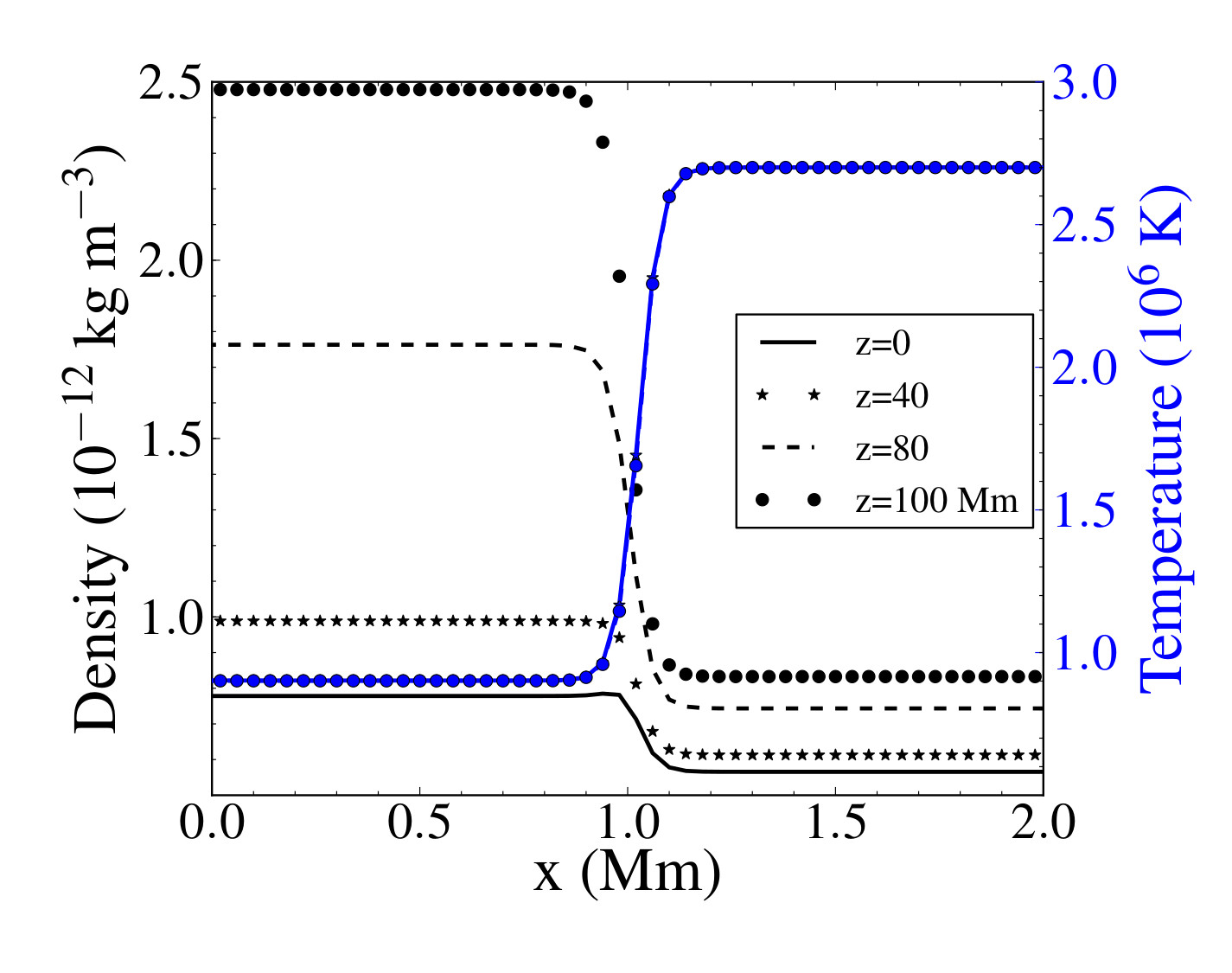}
\end{center}
\caption{Radial profile of the initial density (with black) and temperature (with blue) for our gravitationally stratified, cylindrical flux tubes  at different heights, right before the introduction of the driver. The apex is located at $z = 0$ and the footpoint at $z = 100$ Mm. $x = 0$ is the centre of the loop at $t = 0$.}\label{fig:1prof}
\end{figure}

Gravity varies sinusoidally along the flux tube, corresponding to a semi-circular loop with major radius $L \pi^{-1}$, and takes a zero value at the loop apex ($z=0$) and a maximum absolute value at the footpoints ($z=\pm 100$ Mm). Due to the hydrostatic equilibrium we have stratification of pressure and temperature along the loop. To counteract the initial pressure imbalance at the loop boundary caused by the gravitational stratification, the magnetic field is restructuring inside the flux tube, once we let the system reach a semi-equillibrium state after one period. By the end of the relaxation period, neither of temperature, pressure, nor density deviate significantly from their initial state, as in \citet{karampelas2019}. The radial profiles of density and temperature at several positions along the z-axis are shown in Fig. \ref{fig:1prof}.

\subsection{Boundary conditions and driver}

Our tubes are driven from the footpoint ($z=100$ Mm), using a continuous, monoperiodic `dipole-like' driver \citep{karampelas2017}, inspired by that used by \citet{pascoe2010}. The period of the driver is $P\simeq 2L/c_k$, coinciding with the corresponding fundamental eigenfrequency for our gravitationally stratified flux tube \citep{edwin1983wave, andries2005A&A430.1109A}. For a kink speed of $c_k = 1167$ km s$^{-1}$ we obtain a period of $P = 171$ s. The driver at the bottom boundary has a uniform and time varying velocity inside the loop,
\begin{equation}
\lbrace \upsilon_x,\upsilon_y \rbrace=\lbrace \upsilon(t),0 \rbrace = \left\lbrace \upsilon_0 \cos(\dfrac{2\pi t}{P}),0 \right\rbrace ,\label{eq:2}
\end{equation}
where $\upsilon_0$ (in km s$^{-1}$) is the peak velocity amplitude. Outside the loop, the velocity follows the relation
\begin{equation}
\lbrace \upsilon_x,\upsilon_y \rbrace = \upsilon(t)R^2 \left\lbrace  \frac{(x-\alpha(t))^2-y^2}{((x-\alpha(t))^2+y^2)^2},\frac{2(x-\alpha(t))y}{((x-\alpha(t))^2+y^2)^2} \right\rbrace ,\label{eq:3}
\end{equation}
where $\alpha(t) = \upsilon_0 \, (0.5\,P/\pi)\, \sin(2\pi t/P)$ is a function that recentres the driver, tracking the footpoint. A transition region following the density profile exists between the two areas, in order to avoid any numerical instabilities due to jumps in the velocity. 

We study five different models (\lq D1\rq, \lq D2\rq, \lq D4\rq, \lq D6\rq, and \lq D8\rq,), each for a different corresponding value of $\upsilon_0=1,\, 2, \, 4, \, 6$ and $8$ km s$^{-1}$. 

At the bottom boundary, we also keep the velocity component parallel to the $z$-axis ($v_z$) antisymmetric ($z=100$ Mm) and we extrapolate the values for density and pressure, using the equations for hydrostatic equilibrium. A zero normal gradient condition is used to extrapolate the values of each magnetic field component through the bottom boundary. Studying the fundamental standing kink mode for an oscillating flux tube allows us to take advantage of the inherent symmetries of this mode, as well as the symmetric nature of our driver. In the top boundary at $z=0$, the location of the loop apex, we kept $v_z$, $B_x$, and $B_y$ antisymmetric in the $x-y$ plane at the apex. The rest of the quantities on that boundary are defined as symmetric. Thus, only half the loop is simulated along the loop axis. All side boundaries are set to outflow (Neumann-type, zero-gradient) conditions for all variables.

\subsection{Numerical method and grid}

The 3D ideal MHD problem is solved using the PLUTO code \citep{mignonePLUTO2018}, with the extended GLM method from \citet{dedner2002} keeping the solenoidal constraint on the magnetic field. We use the finite volume piecewise parabolic method (PPM) and the characteristic tracing method for the timestep. The domain dimensions are $(x,y,z) = (16,6,100)$ Mm for models D1, D2 and D4, and $(x,y,z) = (16,10,100)$ Mm for models D6 and D8. The latter was chosen in order to properly resolve the expanded turbulent loop cross-section from the stronger drivers. We have cell dimensions of $40 \times 40 \times 1562.5$ km for all models. The resolution is higher in the $x-y$ plane, to better resolve the small-scale structures that appear in the loop cross section. The density scale height of our setup is $28$ Mm, and is close to the value of the loop major radius $L \pi^{-1}\sim 64$ Mm. Alongside the lack of radiation or thermal conduction, this allows for a coarser resolution on the $z$-axis, that can still sufficiently model the density stratification along the flux tube. In all of our models, we also have the inevitable numerical dissipation effects, which lead to values of effective resistivity and viscosity many orders of magnitude larger than the expected ones in the solar corona. 

\subsection{Forward modelling}

We use the FoMo code \citep{fomo2016} to create synthetic images and compare to real observations. We focus on the use of the 
Fe XII $193$ \AA \,emission line, with maximum formation temperature of $\log T = 6.19$, in order to create spectroscopic data of our flux tubes. The $193$ \AA\,line tracks the warmer flux tube boundary layer \citep{antolin2016,antolin2017} and is better suited for our studies of driven oscillations with turbulent cross-sections \citep{karampelas2019}. The same analysis has been performed for the Fe IX $171$ \AA \,line. This line has a maximum formation temperature of $\log T = 5.93$, and is more sensitive to the colder plasma from the loop interior. We also use the \lq Imaging\rq functionality of \lq fomo-c\rq from FoMo, in order to create emission images for the corresponding AIA channels.

We create time-distance maps of the loop apex for each model, by placing a slit at LOS angles equal to $0^\circ$ (perpendicular to the axis of the oscillation, or $x$ axis), $45^\circ$ and $90^\circ$ (parallel to the axis of the oscillation).To better compare to the observations of decayless coronal loop oscillations, we degrade the original spatial resolution of the AIA synthetic images to that of SDO/AIA ($ 0''.6 $). Then we resample the data to match the pixel size of the target instrument. A similar process is used for the spectroscopic data, where we try to mimic Hinode/EIS by taking a plate-scale of $1''$ and a Gaussian PSF of $3''$ FWHM \citep{antolin2017,mingzhe2019}, while also adding poisson distributed photon noise. For the spectroscopic data we also reduce the initial spectral resolution from $3$ km s$^{-1}$ to $36$ km s$^{-1}$ for the degraded spectrum. The temporal resolution is kept at $\sim 11$ s, which is very close to both that of AIA and EIS instruments. 

\section{Results}

We drive our loops for fifteen cycles. As in \citet{karampelas2017}, the first waves to reach the apex ($z=0$) are the azimuthal Alfv\'{e}n waves at the boundary layer of our tube, thanks to their higher propagation speed, followed by the propagating kink waves. Because of the symmetry at the apex, the driver induced waves lead to the initiation of a standing oscillation resembling the fundamental kink mode for gravitationally stratified loops \citep{andries2005A&A430.1109A,magyar2015,karampelas2019}. 

\begin{figure}[h!]
\begin{center}
\includegraphics[trim={0cm 0.25cm 0cm 0cm},clip,scale=1]{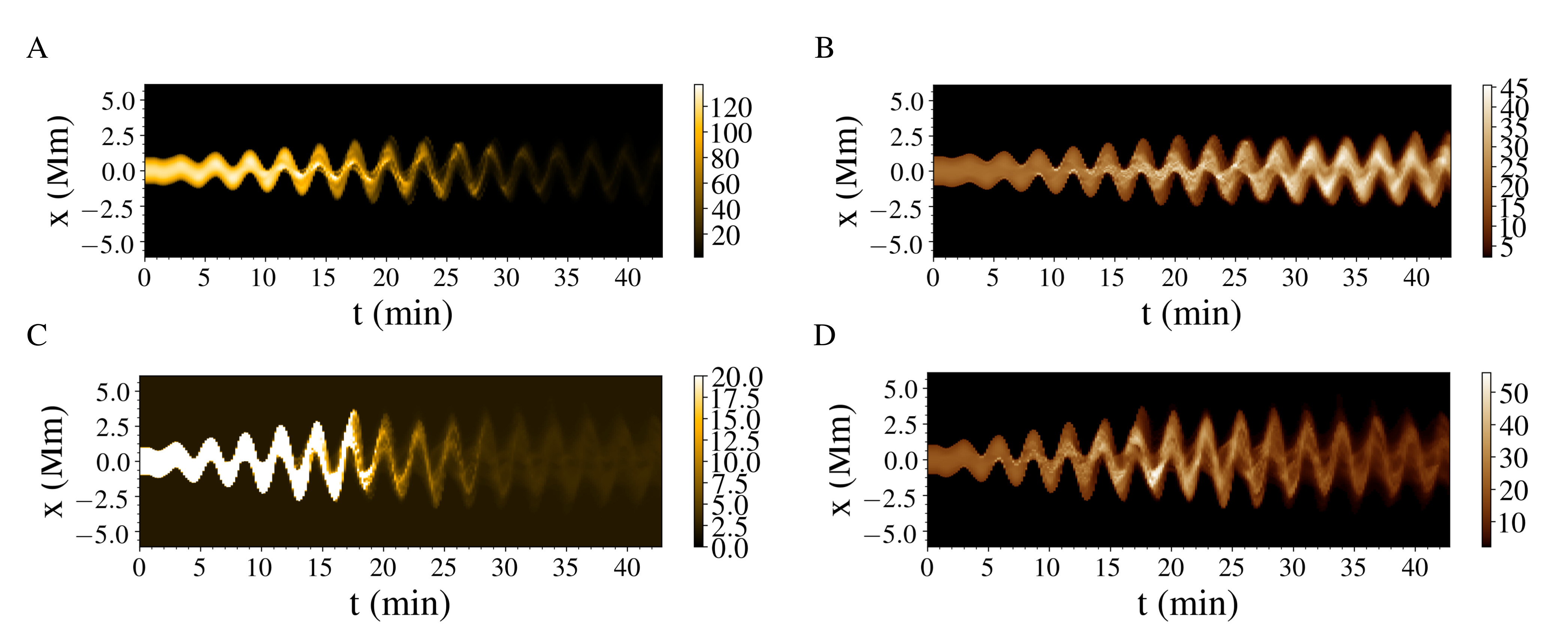}
\end{center}
\caption{Synthetic time-distance maps for the D2 model in the Fe IX $171$ \AA \,\textbf{(A)} and Fe XII $193$ \AA \, \textbf{(B)} and for the D4 model in the Fe IX $171$ \AA  \textbf{(C)} and Fe XII $193$ \AA \,\textbf{(D)} lines at the apex. The time distance maps show the emission (in $ergs\ cm^{-2} s^{-1} sr^{-1}$) for the numerical resolution panels of our model. For the D4 model, the intensity in the $171$ \AA \,line is saturated, in order to clearly depict the drop in the emission. The LOS angle is $0^\circ$.}\label{fig:1pre}
\end{figure}

Once the standing mode is initiated, we have the gradual development of the KH instability and the expansion of the flux tube cross-section \citep{karampelas2017}. The values of the effective numerical dissipation coefficients are low enough not to hinder the development of the instability \citep[see][]{karampelas2019}. By performing forward modeling for our setups, we construct time distance maps of our oscillations at the apex, for the different spectral lines. Two examples of these forward modeling time-distance maps are presented in Fig. \ref{fig:1pre}. There we see the apex displacement over time for the D2 and D4 model in the $171$ \AA \,and $193$ \AA \, AIA channels. As expected for loops with turbulent cross-sections \citep{karampelas2019}, the development of the KH instability and the spatially extended TWIKH rolls lead to extensive mixing of the loop with the surrounding plasma. For our setup of a cold loop embedded in a warmer corona, this leads to a big emission drop in the Fe IX $171$ \AA \,line.


By saturating the image of the $171$ \AA \,emission line for model D4, we can see that this drop occurs faster for stronger drivers. A smaller drop is also observed in the $193$ \AA \,line, for stronger drivers, and is caused by the drop of the average density of the emitting plasma, and its mixing with the surrounding plasma \citep{karampelas2017}, due to the spatially extended TWIKH rolls. In coronal observations we expect to mostly see the cooling stage of loops \citep{ViallKlimchuk2012ApJ}. The loops observed performing these decayless observations are expected to be driven for long periods of time, while cooling from higher initial temperatures. Hence, the observed decayless regime should correspond to the later stages of our simulations, where we observe the mixing induced emission drop.

\begin{figure}[h!]
\begin{center}
\includegraphics[trim={0cm 0.5cm 0cm 0cm},clip,scale=1]{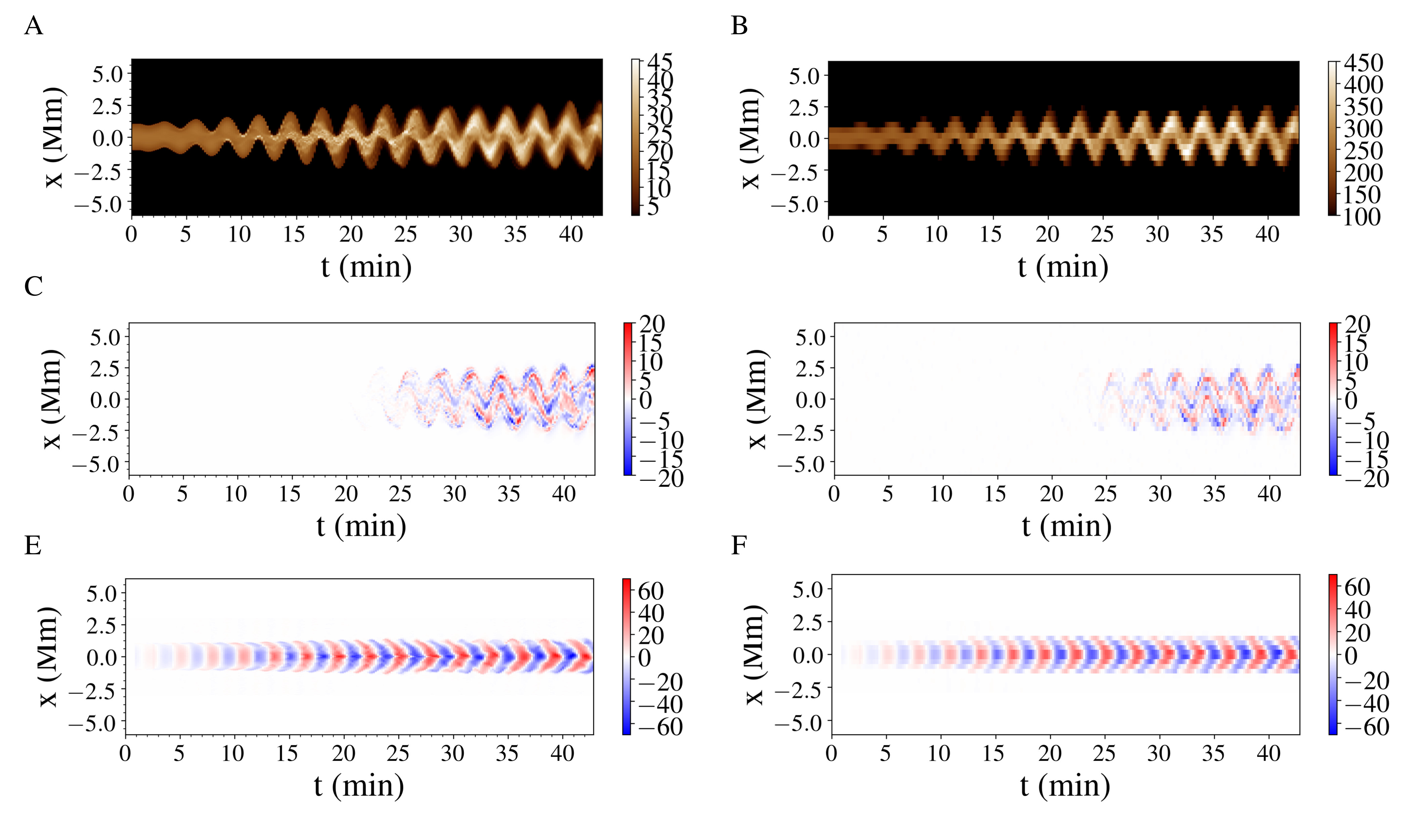}
\end{center}
\caption{Forward modeling results for the D2 model in the Fe XII $193$ \AA \,line at the apex. The left panels show the images for numerical resolution and the right panels the images for the targeted instruments. \textbf{(A)} and \textbf{(B)} show the emission intensity (in $ergs\ cm^{-2} s^{-1} sr^{-1}$) for numerical and for AIA resolution respectively. \textbf{(C)} and \textbf{(D)} show the Doppler velocity (in km s$^{-1}$) at $0^\circ$, for numerical and for EIS resolution respectively. \textbf{(E)} and \textbf{(F)} is the Doppler velocity (in km s$^{-1}$) at $90^\circ$, for numerical and for EIS resolution respectively.}\label{fig:1}
\end{figure}

In Fig. \ref{fig:1} we present additional synthetic images for the D2 model, both at numerical and instrument resolution. In that figure, we see the $193$ \AA \,AIA channel for the emission intensity at $0^\circ$ LOS angle, as well as the Doppler velocities for the Fe XII $193$ \AA \,line at $0^\circ$ and $90^\circ$ angles. We observe the development of emission peaks prior to the development of the TWIKH rolls, due to the excitation of higher order harmonics and the deformation of the flux tube from the combined effect of the inertia and fluting modes \citep{andries2005A&A430.1109A, Ruderman2010PhPl, YuanTVD2016ApJS, antolin2017,terradas2018}. Once the KHI develops, we see the emergence of stronger emission peaks in the $193$ \AA \, channel, stronger Doppler velocities perpendicular to the direction of the oscillation, and the gradual expansion of the resonant flows across the loop cross-section \citep{antolin2015resonant,karampelas2018fd}. The latter is evident in the bottom panels of Fig. \ref{fig:1}, where we look at the loop from the side. These out of phase movements from the TWIKH rolls are present in all of our models and get more intense and spatially extended the stronger each driver is.

\begin{figure}[h!]
\begin{center}
\includegraphics[trim={0cm 0.4cm 0cm 0cm},clip,scale=1]{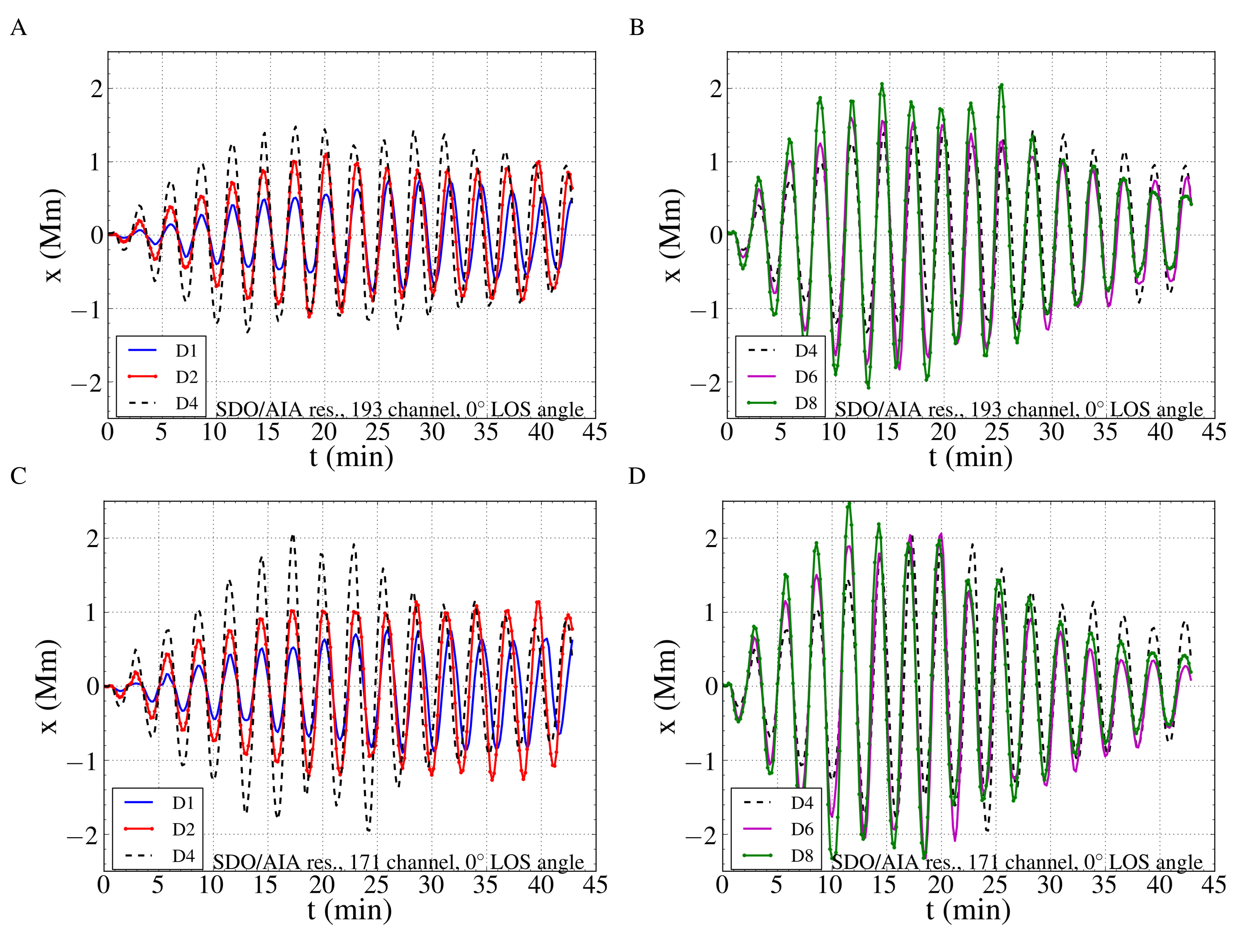}
\end{center}
\caption{Time distance maps of the oscillation amplitude for the five models. The apex displacements were obtained by tracking the loop edge in synthetic emission images of the Fe XII $193$ \AA \,line for \textbf{(A)} and \textbf{(B)} and of the Fe IX $171$ \AA \,line for \textbf{(C)} and \textbf{(D)}, at $0^\circ$}\label{fig:2}
\end{figure}

We track the oscillations in the constructed time-distance maps for our models by setting a threshold value off $I_{thr} = I_{min} + 0.1(I_{max} - I_{min})$, where $I$ is the intensity across the loop on each frame, $I_{max}$ is its maximum value and $I_{min}$ is the background intensity. The point with this intensity is defined as the \lq edge\rq of the loop. A spline interpolation for the position is used to inhibit tracking errors, with the number of interpolation points being half the number of frames in the TD map. We present the displacement of the apex for each loop in Fig. \ref{fig:2}, where panels for all the models are shown at a $0^\circ$ LOS angle. An initial increase of the amplitude is followed by saturation and an eventual drop in the second half of the simulations. This is the stage of the simulations where the decayless regime should correspond to. In the last stages, we see that the D2 and D4 models show similar values of displacement, while the D6 and D8 models show lower amplitudes than the former ones. 

We see that similar profiles are obtained for both emission lines used in the present work. The synthetic images in the $171$ \AA \,line show higher initial amplitudes than those in the $193$ \AA \,line. This is because the loop core has the highest transverse velocities before the development of the KH instability. The loop boundary layer has radially decreasing $\upsilon_x$ velocities, as result of the driver shape that we use at the footpoint, leading to smaller displacements than the loop core, as seen in the two emission lines, each being more sensitive to the corresponding loop region. A similar thing was observed in \citet{antolin2016}. In that work, the loop core showed a higher oscillation amplitude than the loop boundary layer, before the initiation of the KHI. In our models, the constant driving leads to overall stronger displacements, while the deformation of the loop cross-section leads to higher amplitudes in the $193$ \AA \,line for the given temperature distribution. 

In Fig. \ref{fig:2} we get a small phase difference between the D1 model and all the rest. This small change in the  phase between the observed displacement and the driving frequency is a LOS effect and caused by the integration of the emission across TWIKH rolls with varying phases. A similar phase shift was observed for the turbulent layer of oscillating loops performing in \citet{antolin2016}. For the D1 model, the slower development of the KHI leads to a delay in the manifestation of that phase difference. Similar trends have been observed for synthetic images at $45^\circ$, with the difference that the obtained displacements are smaller due to the projection of the oscillation plane on the plane of sky (POS). 

\begin{figure}[h!]
\begin{center}
\includegraphics[trim={0cm 0.3cm 0cm 0cm},clip,scale=1]{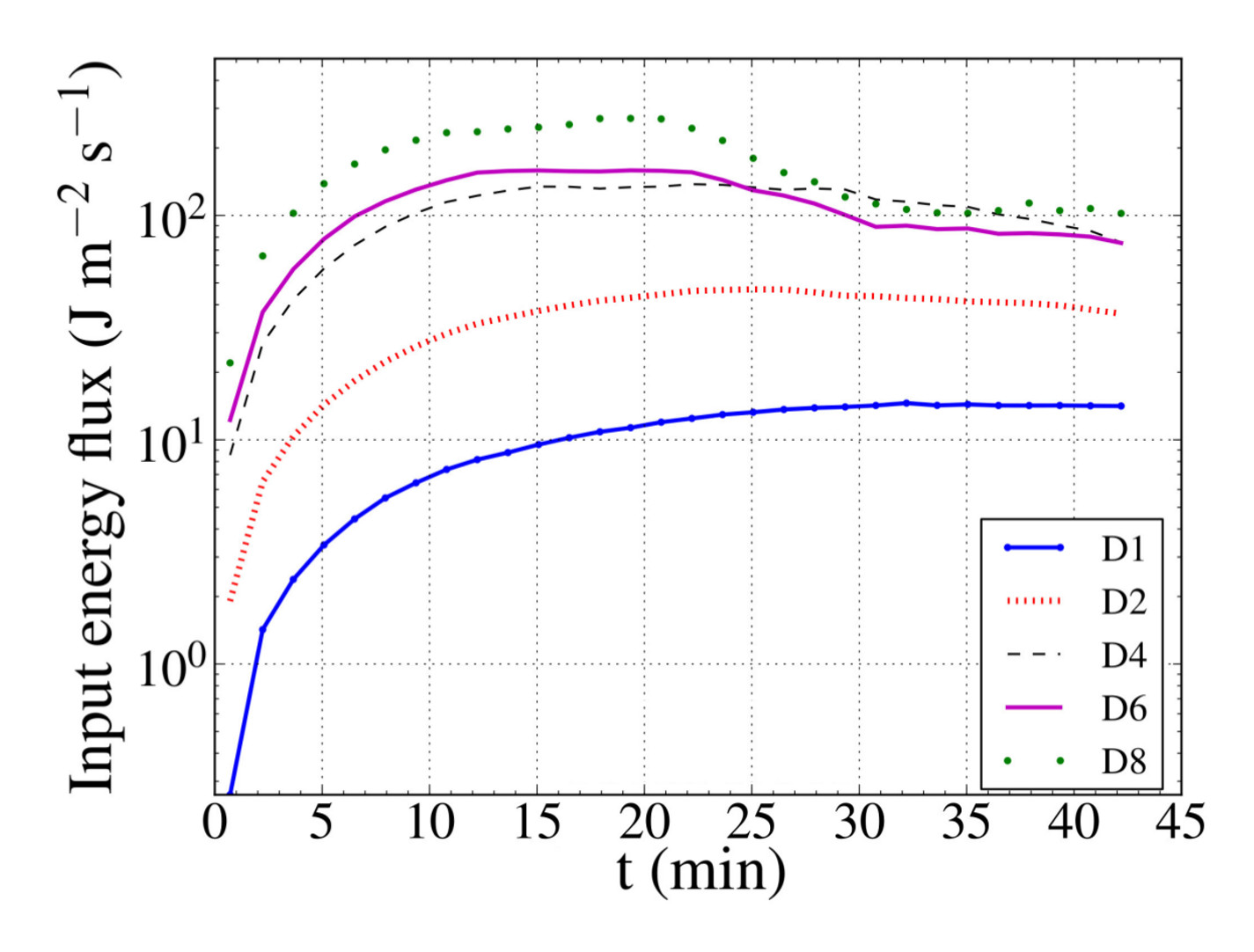}
\end{center}
\caption{Evolution of the average values of the energy fluxes provided by the driver of each model (in W m$^{-2}$) over time.}\label{fig:3}
\end{figure}

In order to determine whether the obtained amplitude drop in Fig. \ref{fig:2} is caused purely by the development of strong azimuthal resonant flows and the KH eddies, we plot in Fig. \ref{fig:3} the input energy flux (J m$^{-2}$ s$^{-1}$) for each driver. The Poynting fluxes are calculated for a surface area of $3\times 3$ Mm$^2$ in the centre of the bottom boundary. This square contains the loop footpoint of each model, and provides $80 \%$ of the total energy input from the spatially localized driver. The curves in Fig. \ref{fig:3} represent the time evolution of the average values of the fluxes.

We can see that the efficiency of the drivers change over time, with the strongest changes happening for the stronger drivers. These variations with time are the result of the boundary conditions employed here. The zero normal gradient condition for the magnetic field allowed it to freely evolve with time, instead of fixing its value at the footpoint. This is affecting the Poynting flux from the driver in the bottom boundary, causing the evolution of the energy input, similarly to \citet{mingzhe2019} and \citet{karampelas2019}. In the scope of this study, we do not focus on creating a realistic model for a driver, but we focus on the effects of the driver strength on the synthetic data. Therefore, tracking the evolution of the driver efficiency is sufficient within the context of the current study. The initial increase in the energy flux is followed by a drop, which is again more pronounced in the models with the stronger drivers. By comparing the curves for the models D2 and D4 with the synthetic images of Fig. \ref{fig:1pre}, we see place that this drop takes place when the loops become turbulent due to the KHI.

Focusing at the last 6 cycles of the the simulation in Fig. \ref{fig:3}, we see that D4 shows a stronger input than D6, while still remaining weaker than D8. D2 on the other hand remains weaker from all the aforementioned ones, providing almost half or less as much input flux. This is in contrast to the results of Fig. \ref{fig:2}, where the oscillation amplitude of the D2 model was comparable to that of the D4 model, and higher than the respective amplitudes of models D6 and D8. This shows that the previously obtained amplitudes are affected by the development of the out of phase movements of the spatially extended TWIKH rolls. All the drivers provide fluxes within the range of the radiative losses of the non-active region corona \citep{withbroenoyes1977ara}, with the exception of D1, which is significantly weaker than the rest. 

\begin{figure}[h!]
\begin{center}
\includegraphics[trim={0cm 0.15cm 0cm 0cm},clip,scale=1]{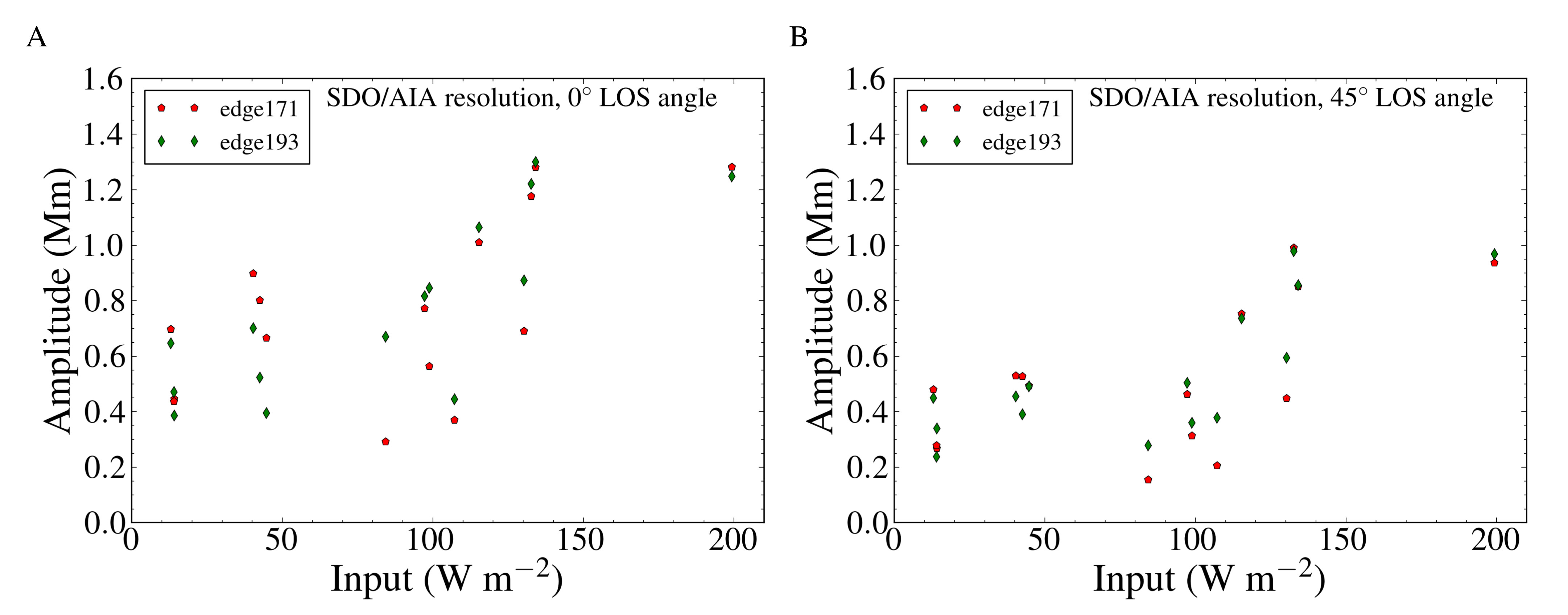}
\end{center}
\caption{Diagrams of the average oscillation amplitudes as a function of the average energy input for each model. In  \textbf{A} we have a $0^\circ$ LOS angle and in \textbf{(B)} we have a $45^\circ$ LOS angle. Data from the Fe XII $193$ \AA \,line and the Fe IX $171$ \AA \,line were used. The loops displacements for the two lines were obtained by tracking their edge in the corresponding synthetic emission images.}\label{fig:4}
\end{figure}

After calculating the input energy flux and the observed loop displacements, we plot amplitude-flux diagrams in Fig. \ref{fig:4}. For these diagrams, we took the average of the (absolute values) of the maximum displacements and fluxes, for each model, at time intervals $19.95-28.5$ minutes, $28.5-34.2$ minutes and $34.2-42.75$ minutes. The diagrams contain the amplitudes traced from AIA-resolution synthetic images of $171$ and $193$ AIA channels at both $0^\circ$ and $45^\circ$ LOS angle. Taking this and the results from Fig. \ref{fig:2} into account, we explain the differences between the  $171$ and $193$ \AA \,AIA channels as LOS effects of our models. The highest values of the amplitudes are obtained for the highest driver input fluxes. However, a clear correlation is absent for fluxes $\leq 110$ W m$^{-2}$. Instead, we see a relatively uniform distribution of amplitudes among the different fluxes in that region. The corresponding amplitudes are below $1$ Mm as we look perpendicular to the oscillation direction ($0^\circ$.) At a LOS angle of $45^\circ$, the value of these amplitudes is between $0.15$ and $0.55$ Mm, placing them within the range of the observed decayless oscillations \citep{nistico2013,anfinogentov2015}. 

Finally, we want to calculate the observed kinetic energy flux (W m$^{-2}$) across the line of sight, for a specific spectral line. Based on the work of \citep{antolin2017}, we calculate the Doppler energy flux as:
\begin{equation}
F_{obs} = \dfrac{1}{2} \int_{\bot LOS} <\rho_{\lambda}(l,t)>  \upsilon_{Dop,\lambda}(l,t)^2 \dfrac{c_k}{L_{\bot LOS}} dl_{\bot},\label{eq:4}
\end{equation}
and the non-thermal energy flux as:
\begin{equation}
F_{obs} = \dfrac{1}{2} \int_{\bot LOS} <\rho_{\lambda}(l,t)> \left( \xi_{\lambda}(l,t)^2 - \xi_{\lambda,th}(l,t)^2 \right) \dfrac{c_k}{L_{\bot LOS}} dl_{\bot}.\label{eq:5}
\end{equation}
With $c_k = 1167$ km s$^{-1}$ we denote the kink speed. $L_{\bot LOS}$ is the length of the domain across the LOS, $\upsilon_{Dop,\lambda}$ the Doppler velocity (in km s$^{-1}$) for the spectral line, $\xi_{\lambda}$ the corresponding Doppler width in km s$^{-1}$, and the thermal width (in km s$^{-1}$):
\begin{equation}
\xi_{\lambda,th} = \dfrac{c}{\lambda_0}\sqrt{\dfrac{T_{\lambda}k_B}{\mu m_p}}.\label{eq:6}
\end{equation}
With $c$ we denote the speed of light, $\lambda_0$ the wavelength at rest for the spectral line, $k_B$ the Boltzmann constant, $m_p$ the proton mass and with $\mu$ the atomic weight in proton masses of the emitting element. The emitting temperature $T_{\lambda}$ is taken as an approximation from the $\xi_{\lambda,th}$ in the beginning of the simulations, where the velocities along the LOS are practically zero. The quantity $<\rho_{\lambda}(l,t)>$ is based on the emissivity-weighted average density from \citet{antolin2017}:
\begin{equation}
<\rho_{\lambda}(l,t)> = \dfrac{\int_{// LOS} \rho \epsilon(l,t) ds}{\left( \int_{// LOS} \epsilon(l,t) ds \right) \mid_{max}},\label{eq:7}
\end{equation}
where we instead divide the integral of density times emissivity with the maximum value of the integrated emissivity along the LOS at each frame. The fluxes from all setups are then normalized to the domain dimensions of models D6 and D8 for all setups, to make direct comparisons possible between the different models.

\begin{figure}[h!]
\begin{center}
\includegraphics[trim={0cm 0.4cm 0cm 0cm},clip,scale=1]{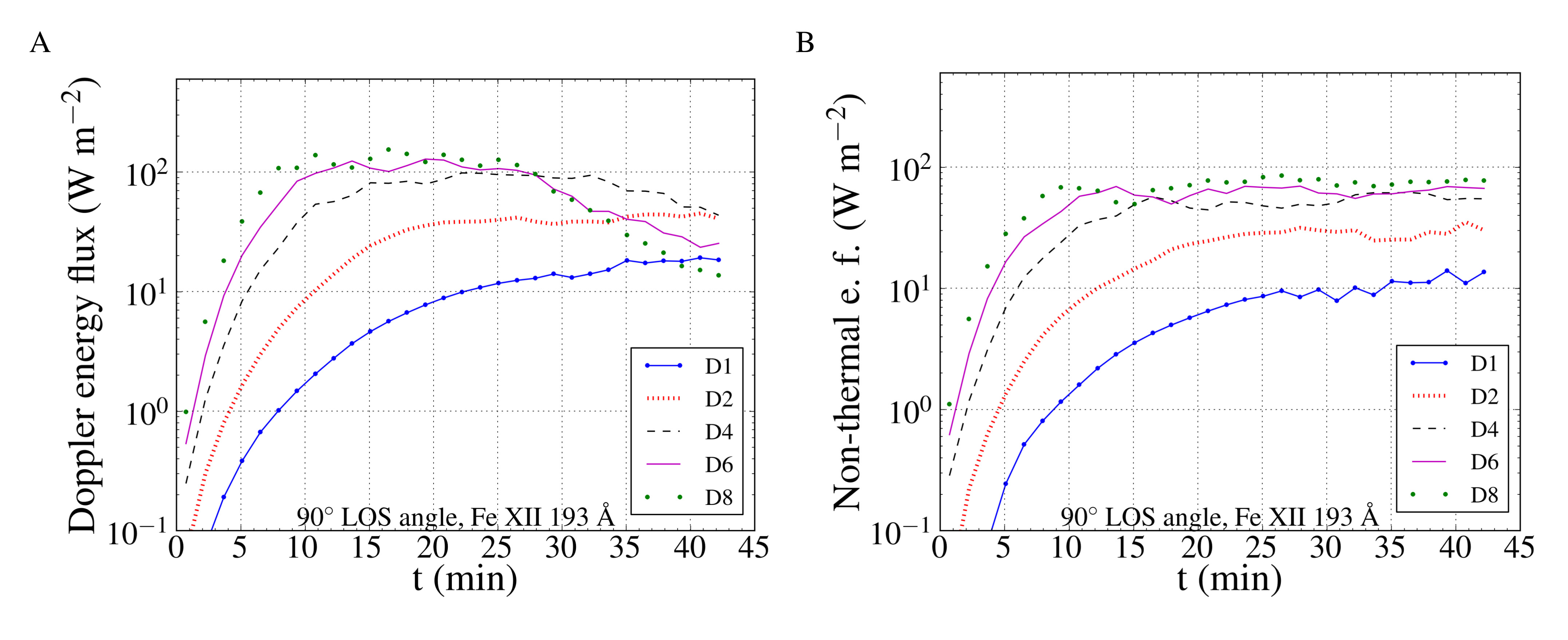}
\end{center}
\caption{Average values of the Doppler \textbf{(A)} and non-thermal \textbf{(B)} energy fluxes for a LOS angle of $90^\circ$, based on synthetic spectroscopic data from the Fe XII $193$ \AA \,line.}\label{fig:5}
\end{figure}

We plot our results for the Fe XII $193$ \AA \,spectral line in Fig. \ref{fig:5}, following Hinode/EIS spatial, spectral and temporal resolution. For a LOS angle of $90^\circ$, we observe an initial increase in the energy flux estimated by the Doppler velocities, which persisted for the duration of the simulations for models D1 and D2. For the models D4, D6 and D8 we observe a drop in the estimated Doppler energy flux, in the last cycles of the simulations. This drop appears once the turbulent loop cross-section shows very strong out of phase flows, due to the TWIKH rolls. These azimuthal flows do not register in the Doppler velocities at this angle, leading to the aforementioned drop. A similar behavior was observed in \citet{antolin2017} for a non-driven transverse oscillation. We also see a correlation with the drop of the input energy flux, as observed for the the models D4, D6 and D8. However, a similar correlation is not observed for the D1 and D2 models. This shows that the observed drop in the Doppler energy flux is the combined result of the input drop and the loop deformation by the TWIKH rolls.

The diagram of the non-thermal energy flux in Fig. \ref{fig:5} shows a saturation once the KHI fully deforms the loop cross-section. This agrees with the results of \citet{antolin2017}, where a similar saturated domain was obtained after the manifestation of the KHI. For the models D4, D6 and D8, this saturation takes place before the drop in the Doppler energy flux. The small increase of the non-thermal energy flux observed in the later parts of the simulation, does not fully compensate for the drop in the Doppler energy flux. This shows that the aforementioned Doppler flux follows closely the  input energy flux, in the case of continuously driven oscillating loops. We also observe that the non-thermal energy fluxes are increasing for stronger drivers. This suggests that the observed correlation between line widths and Doppler shifts reported by \citet{mcintosh2012} can correspond to a continuous input of transverse MHD waves, supporting their hypothesis. However, the use of a single loop for each driver excludes the LOS effects of integrating over many structures, and cannot give a safe statistical result to be compared with the observations \citep{mcintosh2012}.

\begin{figure}[h!]
\begin{center}
\includegraphics[trim={0cm 0.4cm 0cm 0cm},clip,scale=1]{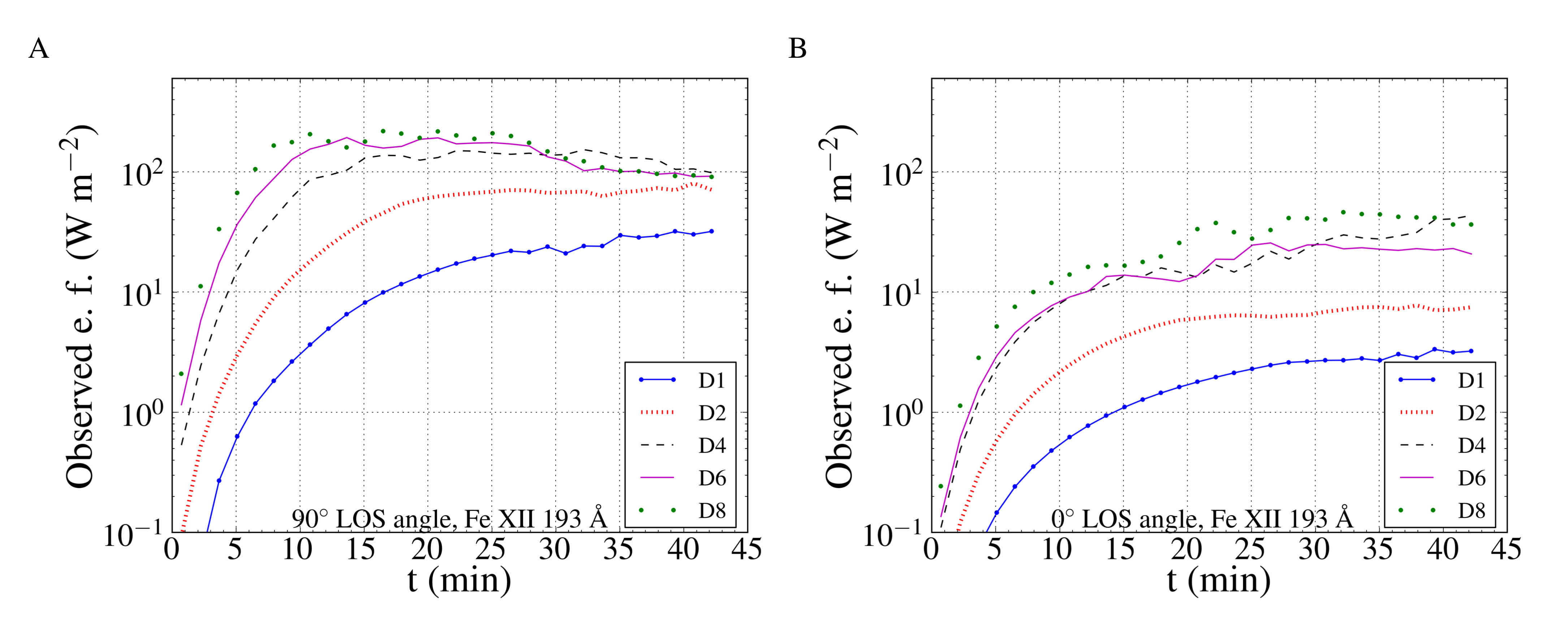}
\end{center}
\caption{Average values of the total observed (Doppler and non-thermal) energy fluxes, based on synthetic spectroscopic data from the Fe XII $193$ \AA \,line. A LOS angle of  $90^\circ$ \textbf{(A)} and  $0^\circ$ \textbf{(B)} was considered.}\label{fig:6}
\end{figure}

We see a disagreement between the total observed (Doppler $+$ non-thermal) energy flux, for a LOS angle of $90^\circ$, and the input in Fig.\ref{fig:3} for models D1 and D2. The former, as seen in Fig. \ref{fig:6} is overestimated with respect to the input energy flux during the later part of the respective simulations. This comes in contrast with the results of \citet{antolin2017}, where the estimated total energy flux was lower than the available energy flux calculated directly from the simulation data. This disagreement is due to the inaccuracies introduced through the method of estimating the density and temperature of the emitting plasma from the synthetic data. The latter appears in the correction for the thermal terms in the non-thermal energy fluxes. Additional uncertainty can be caused by the angle of observation. As seen in Fig. \ref{fig:6}, the total observed energy for a LOS angle of $0^\circ$ follows the development of the TWIKH rolls and leads to an underestimation of the energy flux.

\begin{figure}[h!]
\begin{center}
\includegraphics[trim={0cm 0cm 0cm 0cm},clip,scale=1.2]{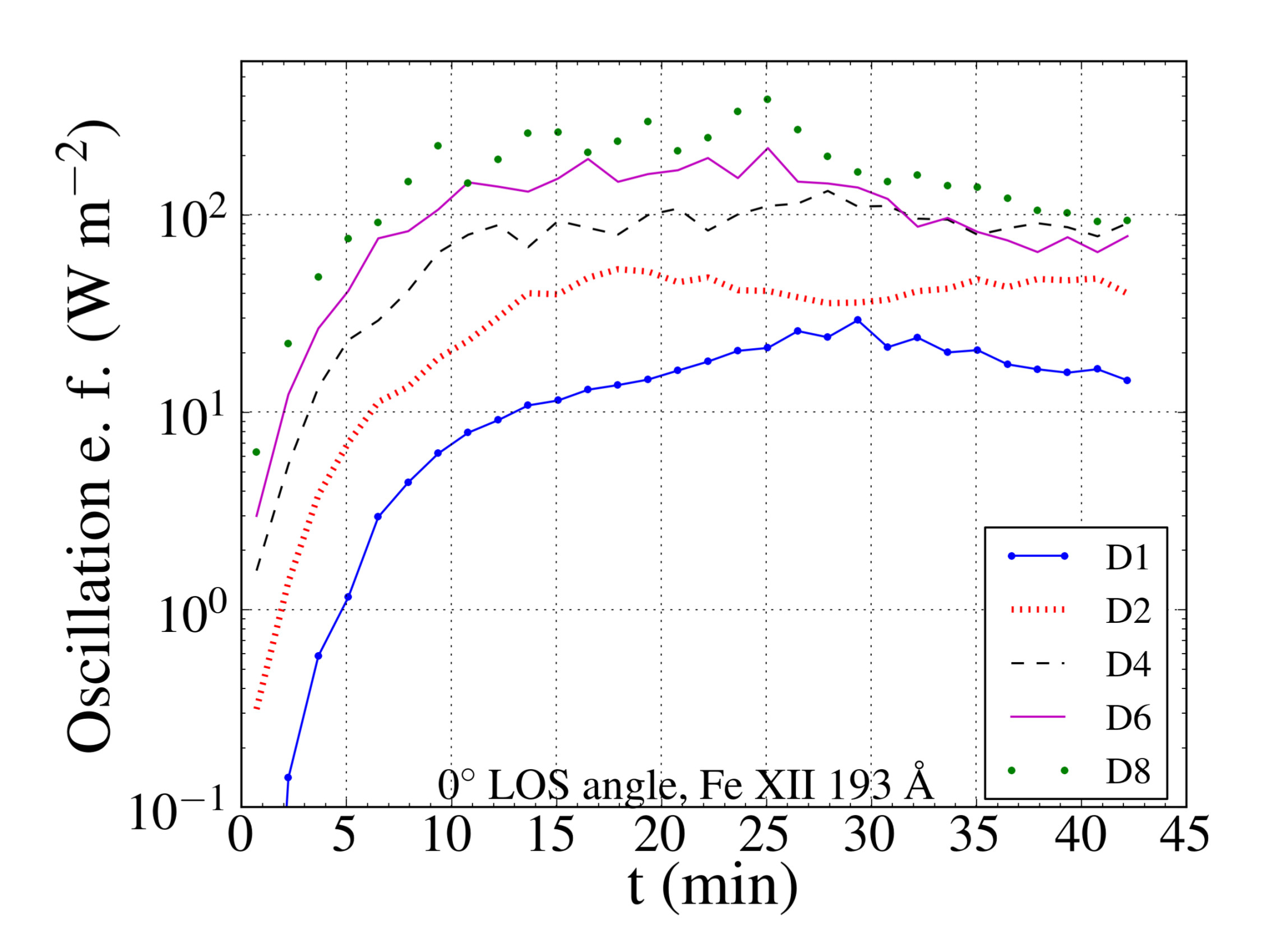}
\end{center}
\caption{Average values of the total observed (Doppler and non-thermal) energy fluxes with the additional term by tracking the oscillation on the POS. The values are based on the synthetic images for the Fe XII $193$ \AA \,line and the $193$ AIA channel.}\label{fig:7}
\end{figure}

The uncertainty revolving around the LOS angle can be bypassed by a combined approach of multiple observations. Once a small angle of observation is identified by combining the energy diagrams and the spectra, like those in Fig. \ref{fig:2}, we can then approximate the total energy fluxes. We achieve this by adding an additional term to the sum of the Doppler and non-thermal energy fluxes. This additional term is derived from eq. \ref{eq:4}, by replacing the Doppler velocities with the the velocity derived from the displacement of the oscillating loop, like in the time-distance maps of Fig. \ref{fig:2}. The results of these calculation are presented in Fig. \ref{fig:7}, and show a better agreement with the input energy than the underestimated fluxes in panel (B) of Fig. \ref{fig:6}. 

\section{Discussion}
In the present work we aimed to study the effects of different transverse drivers in the dynamics and energetics of decayless
oscillations in coronal loops, through the use of synthetic images. We modeled the decayless oscillations as the standing waves from continuously footpoint-driven straight flux tubes, in the presence of gravity. We performed MHD simulations for footpoint drivers of different strengths and performed forward modeling to our data, constructing time-distance maps at the loop apex. We then degraded the spatial and spectral resolution of our signal, targeting different instruments, namely the SDO/AIA and Hinode/EIS. From the created acquired synthetic images for the $193$ \AA \,and the $171$ \AA \,lines we studied the oscillation amplitudes in the plane of sky (POS) at the apex, generated by the five different drivers. To do so, we tracked the \lq edge\rq of the loop from the emission images,
which is defined by threshold value for the intensity, set slightly higher than the background intensity. 

In Fig. \ref{fig:1pre} we observe the drop of the emission in synthetic images for both lines employed here. This drop is the result of the spatially extended TWIKH rolls, expanding the loop cross-section and causing mixing with the surrounding plasma. This drop in the emission is stronger and occurs faster for stronger drivers, as we can see in the example of Fig. \ref{fig:1pre} for the D2 and D4 models. In observations we mostly expect to observe loops during their cooling stage \citep{ViallKlimchuk2012ApJ}, starting from unknown higher temperatures of formation. For decayless oscillations, we therefore expect them to be already in a turbulent state, because of the continuous driving. Any change in the strength of the driver could then lead to sudden changes in the observed emission in various lines, depending on the temperature gradient between the loop and the surrounding plasma, as well as the temperature of said coronal plasma.

In Fig. \ref{fig:2} we see that the final values of the amplitudes are comparable between the different drivers. We are mostly interested in the second half of the simulations, during which the KHI has deformed the cross-section of each loop. This is because we expect the loops performing decayless oscillations to be already in a turbulent state. During that phase, we can observe higher oscillation amplitudes from weaker drivers. This becomes obvious by comparing these time distance maps and the input energy flux diagrams of Fig. \ref{fig:3} for the D4 and D8 models. The D4 model showed bigger displacements towards the end of the simulations than D8 model, despite having a less efficient driver at that stage of the simulations. In addition to that, the D2 model showed very similar amplitudes to the D4, D6 and D8 models, despite providing less than half the energy flux of these drivers. This result is a clear indication of the importance of the out of phase flows, developed by the TWIKH rolls, in the characteristics of the observed oscillations.

Following the results of Fig. \ref{fig:5}, we observe a drop in the Doppler energy flux for the models D4, D6 and D8. This drop is primarily related to the respective drop of the input energy flux, and secondarily to the development of the TWIKH rolls, since the drop in the input energy flux of models D1 and D2 does is not followed by a drop in their Doppler flux. This is related to the less deformed loop cross-sections expected for weaker drivers in continuously driven oscillations of coronal loops \citep{karampelas2018fd}. The connection between the Doppler energy flux and the resonant flows has also been established in \citet{antolin2017}, for a small-amplitude non-driven transverse oscillation of coronal loop.

From the diagram of the non-thermal energy flux over time in Fig. \ref{fig:5}, we observe the saturation of the energy flux, attributed to the turbulent motions present across the loop cross-section. We also observe a positive correlation between the average values of the saturated non-thermal energy fluxes and the driver efficiency from Fig. \ref{fig:3}. This suggests that 
a continuous input of transverse MHD waves could lead to the observed correlation between line widths and Doppler shifts reported by \citet{mcintosh2012}. However, the small number of measurements (one observation per loop per driver) does not allow us to acquire safe statistical results. Furthermore, our study has excluded the LOS effects of integrating over many structures, which prevents us from making a direct comparison with the observations \citep{mcintosh2012}.

By taking into account the results of Fig. \ref{fig:7}, we see that the sum of the non-thermal and Doppler energy flux can lead to different values, depending on the angle of observation and the methods used in the calculations. By adding an additional term, derived by estimating the oscillation velocity from the time-distance maps, we can compensate for the effects of the observation angle when calculating the energy fluxes. 

We need to stress here that the observed saturation in the amplitudes is not caused by numerical dissipation, but is instead  the combined result of the strong out of phase flows generated by the TWIKH rolls and the effects of plasma mixing in the emission. Scaling tests with setups of higher resolution, as well as past results \citep{mingzhe2019,karampelas2019} have revealed a similar saturation in the oscillation amplitudes. In \citet{karampelas2019}, we have seen that the input energy from our drivers turns into kinetic energy (strong out of phase flows), magnetic energy and internal energy of the plasma (wave heating). The derived spectra in that study reveal a turbulent profile, represented by the inertial range. This inertial range is connected to the existence of smaller scales and is present as long as the KHI is not suppressed by the very large dissipation parameters. The development of these smaller scales could be observed in the Doppler velocities and non-thermal line-widths from spectroscopic observations. These observations, however, would be dependent upon the angle of observation, the instrumental resolution, as well as the assumptions used for the initial temperature and density profile.

Finally, from Fig. \ref{fig:4} we conclude that we cannot obtain a clear correlation between the driver input and the observed oscillation amplitudes just from the synthetic imaging data. In our setups, the amplitudes between $0.15$ and $0.55$ Mm were evenly distributed over a wide range of input fluxes, up to $110$ W m$^{-2}$. Small amplitude oscillations can potentially be hiding enough energy to sustain the non-active region corona ($\sim 100$ W m$^{-2}$) \citep{withbroenoyes1977ara}. The development of spatially extended TWIKH rolls can mask its observational signatures, by causing out of phase movements of the loop plasma and affecting the emission at various lines. This is relevant when explaining the observed decayless oscillations \citep{nistico2013,anfinogentov2015} as the result of continuous footpoint driving of a coronal loop, and stresses the need for combined instrument observations.

To summarize our results, the mixing of plasma caused by the developed KHI affects the emission in synthetic images, leading to a drop once the loop cross-section becomes turbulent. The developed TWIKH rolls show strong out of phase flows, which lead to the saturation of the oscillation amplitudes for each model. For different angles, we obtain oscillation amplitudes near the observed ones, which do not show a clear correlation with the input energy from the driver. These observed amplitudes can potentially carry enough energy to sustain the radiative losses from the Quiet Sun. A better correlation can be obtained between the input energy and the spectroscopic results from the Doppler energy flux and the non-thermal energy fluxes. All of our results, however, are dependent upon the angle of observations and the approximations used, hinting towards the need for combined spectroscopic and imaging observations.

Apart from our suggestion that small-amplitude decayless oscillations contain enough energy flux to support the QS, there are other ways to introduce additional energy within the context of such observed waves. In \citet {demoortel2012ApJ} it was shown that the wave energy is significantly underestimated by integration of multiple loops along the LOS. In that work, the  kinetic energy estimated by the LOS Doppler velocities was $\sim 5\% - 20\%$ of the energy in the domain. A combination of a kink and Alfv\'{e}n driver \citep{mingzhe2019} can also provide more energy into the system, while retaining a similar profile in AIA synthetic images. Alternatively, the use of a broadband driver could also increase the driver efficiency, for only small deviations from the expected oscillation amplitudes. The detection of such higher harmonics in decayless oscillations by \citet{duckenfield2018} suggests that the latter method should be studied further in future works. 

\section*{Conflict of Interest Statement}

The authors declare that the research was conducted in the absence of any commercial or financial relationships that could be construed as a potential conflict of interest.

\section*{Author Contributions}

K.K. has performed numerical simulations and forward modeling for the models in this project, and set the basic ideas for the analysis of the results. D.J.P. has performed specialized data extraction and analysis of the synthetic images. T.V.D. has set the basic guidelines for this project, consulted during the analysis and supervised the completion of the project. M.G. and P.A. have contributed on the methods of the analysis and have given feedback during the discussion of the results. K.K. has written the first draft of the current paper, received input from all co-authors, and written the final version of the manuscript.

\section*{Funding}
K.K., T.V.D. and D.J.P. and were funded by GOA-2015-014 (KU Leuven). M.G. is supported by the China Scholarship Council (CSC) and the National Natural Science Foundation of China (41674172). P.A. acknowledges funding from his STFC Ernest Rutherford Fellowship (No. ST/R004285/1). This project has received funding from the European Research Council (ERC) under the European Union's Horizon 2020 research and innovation programme (grant agreement No 724326). 

\section*{Acknowledgments}
The computational resources and services used in this work were provided by the VSC (Flemish Supercomputer Center), funded by the Research Foundation Flanders (FWO) and the Flemish Government – department EWI. The results were inspired by discussions at the ISSI-Bern and at ISSI-Beijing meetings.

\bibliographystyle{frontiersinSCNS_ENG_HUMS} 
\bibliography{KKarampelasdecaylesssim}

\section*{Figure captions}


%

\end{document}